# NEW DISCOVERIES IN THE GALACTIC NEIGHBORHOOD THROUGH ADVANCES IN LABORATORY ASTROPHYSICS

Submitted by the

American Astronomical Society Working Group on Laboratory Astrophysics (WGLA)
http://www.aas.org/labastro

Nancy Brickhouse - Harvard-Smithsonian Center for Astrophysics
nbrickhouse@cfa.harvard.edu, 617-495-7438

John Cowan - University of Oklahoma
cowan@nhn.ou.edu, 405-325-3961

Paul Drake - University of Michigan
rpdrake@umich.edu, 734-763-4072

Steven Federman* - University of Toledo
steven.federman@utoledo.edu, 419-530-2652

Gary Ferland - University of Kentucky
gary@pa.uky.edu, 859-257-879

Adam Frank - University of Rochester
afrank@pas.rochester.edu, 585-275-1717

Eric Herbst - Ohio State University
herbst@mps.ohio-state.edu, 614-292-6951

Keith Olive - University of Minnesota
olive@physics.umn.edu, 612-624-7375

Farid Salama* - NASA/Ames Research Center
Farid.Salama@mail.arc.nasa.gov, 650-604-3384

Daniel Wolf Savin - Columbia University
savin@astro.columbia.edu, 1-212-854-4124,

Lucy Ziurys – University of Arizona
lziurys@as.arizona.edu, 520-621-6525

*Co-Editors

## 1. Introduction

As the Galactic Neighborhood (GAN) panel is fully aware, the next decade will see major advances in our understanding of this area of research. To quote from their charge, these advances will occur in studies of "the galactic neighborhood, including the structure and properties of the Milky Way and nearby galaxies, and their stellar populations and evolution, as well as interstellar media and star clusters".

Central to the progress in these areas are the corresponding advances in laboratory astrophysics that are required for fully realizing the GAN scientific opportunities within the decade 2010-2020. Laboratory astrophysics comprises both theoretical and experimental studies of the underlying physics and chemistry that produces the observed astrophysical processes. The 5 areas of laboratory astrophysics that we have identified as relevant to the GAN panel are atomic, molecular, solid matter, plasma, and nuclear physics.

In this white paper, we describe in Section 2 some of the new scientific opportunities and compelling scientific themes that will be enabled by advances in laboratory astrophysics. In Section 3, we provide the scientific context for these opportunities. Section 4 briefly discusses some of the experimental and theoretical advances in laboratory astrophysics required to realize the GAN scientific opportunities of the next decade. As requested in the Call for White Papers, Section 5 presents four central questions and one area with unusual discovery potential. Lastly, we give a short postlude in Section 6.

## 2. New scientific opportunities and compelling scientific themes

The significant progress in our understanding of the Galactic neighborhood over the past decade has led to fundamental questions that offer scientific opportunities for the next one. Spectroscopic measurements at millimeter, sub-millimeter, infrared, visible, and ultraviolet wavelengths with high signal to noise and analyses based on new precise atomic, molecular, and nuclear data are providing new insights into chemical evolution. Computer simulations in combination with interferometric observations are indicating the need to describe the interstellar medium (ISM) as a very dynamic entity. Observations in the ultraviolet and x-ray portions of the spectrum are revealing the physical conditions in gas within the halo, the connections between the dynamic ISM in the disk and flows in the halo, and the relationship with extra-planar gas in other galaxies. Improved data of high quality are also probing the dynamic phenomena occurring at the Galactic Center and how the magnetic fields there are playing a role. Observations in the millimeter and infrared are probing the evolution of molecules in interstellar clouds into solid dust grains, leading to key information on the evolution of the ISM. Probably the most exciting opportunity lies in the field of astrobiology, where interdisciplinary efforts are seeking answers to fundamental chemical and biological questions.

## 3. Scientific context

The origin of the elements is one of the central problems of nuclear astrophysics. For elements lighter than iron, the combination of improved atomic data and refined models of stellar atmospheres has yielded precise abundances with improved accuracy (e.g., Scott et al. 2009). Approximately half of the elements heavier than iron were synthesized in billion-degree explosions characteristic of core-collapse supernova, in the process of rapid neutron capture. Observations of old, metal-poor stars have found *r*-process distributions remarkably similar to that of our Sun, suggesting that the *r*-process mechanism is a unique event, operating in the first stars as it does now, producing a characteristic nucleosynthetic signature. Recent progress came about as a result of new laboratory measurements on transition probabilities (e.g., Den Hartog et al. 2006; Lawler et al. 2008). Many of the other heavy elements were produced by slow neutron capture in low mass AGB stars and in the He- and C-burning shells of massive stars. Nuclear data acquired at stellar temperatures (Heil et al. 2008) and modeling of stellar interiors (The et al. 2007; Heil et al. 2008) have greatly improved our understanding of the *s*-process. Stellar models are usually developed to reproduce meteoritic abundances (Lodders 2003). The next level of our understanding of chemical evolution will come from a combination of high quality astronomical spectra and more precise atomic and nuclear data that will be needed to interpret the observations.

Dynamical phenomena control many facets of the ISM, from the formation and destruction of interstellar clouds to the onset of star formation to flows in the Galactic halo. Supernova blast waves and winds from massive stars provide the necessary energy input. Numerical modeling has reached a stage where direct comparison with astronomical observations is providing keen insight into the processes taking place (e.g., de Avillez & Breitschwerdt 2004). Observations from radio (Gaensler & Slane 2006) to x-rays (Hwang et al. 2004) are beginning to reveal the dynamics of supernova remnants and of their interactions with the ISM. Ions seen in ultraviolet and x-ray spectra of the Galactic halo (e.g., Ganguly et al. 2005; Williams et al. 2007) are providing key information on the processes creating the distribution of charge states. The Galactic fountain is a popular model describing such phenomena (Shapiro & Field 1976; Bregman 1980). Controversy, however, surrounds the interpretation of the x-ray measurements. Are the measurements really probing the halo, or instead the intergalactic medium? Similar phenomena appear to be present in the halos of edge-on spiral galaxies (e.g., Tüllman et al. 2006). Further analyses involving observations and simulations will provide insight into this facet of galactic evolution.

The Galactic Center harbors a supermassive black hole, but it appears to be underluminous compared to active galactic nuclei. Studies of flaring events at x-ray wavelengths are likely to provide insight into this difference (e.g., Porquet et al. 2008). Radio observations reveal the presence of non-thermal emission associated with the filamentary nature of the magnetic field there. Such measurements also provide evidence for massive star formation (e.g., Yusef-Zadeh et al. 2008). Our understanding of the nearest active galactic nucleus requires improvements in our knowledge of atomic and molecular plasmas in extreme conditions.

Over the past 30 years, space-based and ground-based astronomy have shown that the Universe is highly molecular in nature. The discovery of over 140 different chemical compounds in interstellar gas, with the vast majority organic molecules, and the realization that obscuring dust pervades vast regions of the interstellar medium have revealed the complexity of interstellar chemistry. Protogalaxies and the first stars are predicted to have formed from primordial clouds where $H_2$ and HD controlled the cooling and collapse of these clouds. Subsequent stars and planetary systems are known to form out of the most complex molecular environments deep inside cold gas and dust clouds, often obscured by hundreds of visual magnitudes of extinction; therefore, it is inevitable that interstellar chemistry is intimately connected to the origins of life. An understanding of the molecular component of the Universe requires a two-fold approach. First, the chemical compounds, their abundances, and how they are distributed in astronomical sources need to be determined. Second, molecular formation mechanisms including reaction pathways and dynamics need to be understood. Attaining these goals is crucial in guiding future astronomical observatories designed to provide new insight into galactic evolution.

## 4. Required advances in Laboratory Astrophysics

Advances particularly in the areas of atomic, molecular, solid matter, plasma, and nuclear physics will be required for the scientific opportunities described above. Here we briefly discuss some of the relevant research in each of these 5 areas of laboratory astrophysics. Experimental and theoretical advances are required in all these areas for fully realizing the GAN scientific opportunities of the next decade.

### 4.1. Atomic Physics

Analyzing and modeling spectra begin with identifying the observed lines that may be seen in emission or absorption. This requires accurate and complete wavelengths across the electromagnetic spectrum. The next step toward understanding the properties of an observed cosmic source depends on accurate knowledge of the underlying atomic processes producing the observed lines. Oscillator strengths and transition probabilities are critical to a wide variety of temperature and abundance studies from infrared to X-ray wavelengths. Many existing data for the heavier elements are still notoriously unreliable. The current limitations on the atomic data available for mid-Z elements make it difficult to determine the nature of the *r*-process.

Modeling ionization structure requires accurate data on many processes. Photoionized gas requires reliable low temperature dielectronic recombination (DR) and electron ionized gas high temperature DR. Calculating low temperature DR is theoretically challenging and for some systems laboratory measurements are the only way to produce reliable data. For high temperature DR, few benchmark measurements exist for L-shell and M-shell ions. Density-dependent DR rate coefficients are needed for dense plasmas. For decades astrophysicists have had to rely on theoretical photoionization calculations of varying degrees of sophistication. The development of third generation synchrotron light sources has opened up the possibility of measuring photoionization cross sections for many important ions. Charge exchange (CX) recombination with H and He and CX ionization with $H^+$ and $He^+$ have been shown to be important for many systems, but few modern calculations or laboratory measurements exist at the relevant temperatures. Data are also needed for low charge states of elements such as Se and Kr in order to study nucleosynthesis in stars that are the progenitors of planetary nebulae.

Line ratios, a key diagnostic, involve knowledge of collision strengths and related phenomena. Rate coefficients for electron impact excitation approaching 10% accuracy are necessary for the most important line ratio diagnostics yielding temperature, optical depth, density, and abundance. Proton impact excitation is important because ions in hot post-shock material decouple from radiatively cooling electrons and may remain hot enough to produce line emission through collisional impact. Furthermore, electron impact ionization (EII) data are highly suspect. Recommended data derived from the same scant set of measurements and calculations can differ by factors of 2 to 3. Much of the published experimental data include contributions from an unknown metastable fraction in the ion beams used. The recommended EII data have not undergone any significant revision or laboratory benchmarking since around 1990. Little data also exist for three-body recombination, the time reverse of EII, which is important in high density plasmas.

## 4.2. Molecular Physics

High-resolution laboratory spectroscopy is absolutely essential in establishing the identity and abundances of molecules observed in astronomical data. Given the advancements in detector technologies, laboratory measurements need to have a resolving power higher than the astronomical instruments. This is essential to interpret ubiquitous interstellar spectral features such as the IR emission bands ("PAH bands") and the diffuse interstellar bands (DIBs) that represent a reservoir of cosmic organic material and hold the key to our understanding of the molecular universe (Salama 1999). Furthermore, all the main functional groups known to organic chemists have now been observed in interstellar molecules, suggesting that the origin of life may have begun in the gas phase chemistry of interstellar clouds. Laboratory spectroscopy is crucial in making the link between interstellar molecules and simple biological compounds that could seed life. It is also crucial in making the link between interstellar molecules in the gas phase and dust grains in the solid phase, a key phase in our understanding of the evolution of interstellar clouds and of circumstellar mass loss. For molecular data obtained from astronomical observations to be of practical use, accurate assignments of observed spectral features are essential. The problem here is two-fold. First, the transitions of known molecules need to be assigned in these spectra, including higher energy levels and new isotopic species. Second, the spectra of undiscovered species that promise to serve as important new probes of astronomical sources need to be identified.

The spectroscopic study of interstellar molecules, many of which are complex structures that cannot be produced in large abundance in the laboratory, requires the development and application of state-of-the-art ultra-sensitive spectroscopic instruments. Detecting the possible presence of a species, however, is not sufficient since it must be reconciled with other physical properties of the medium. To understand the chemical composition of these environments and to direct future molecular searches in the framework of future astronomical observatories, it is important to untangle the detailed chemical reactions and processes leading to the formation of new molecules in extraterrestrial environments. These data, together with quantum chemical calculations, will establish credible chemical models of interstellar, planetary, and stellar environments that are imperative to predict the existence of distinct molecules in extraterrestrial environments, thus guiding future astronomical searches of hitherto unobserved molecular species.

## 4.3. Solid Materials

In order to properly decipher the mechanisms that occur in interstellar cloud environments, laboratory studies of dust and ices are required (Tielens 2005). Studies of silicate and carbonaceous dust precursor molecules and grains are needed as are studies of the dynamical interaction between dust and its environment (including radiation and gas). Studies of dust and ice provide a clear connection between astronomy within and beyond the solar system. Astronomical observations and supporting laboratory experiments from the X-ray domain to the sub-millimeter regions are of paramount importance for studies of the molecular and dusty universe. Observations at infrared and sub-millimeter wavelengths penetrate the dusty regions and probe the processes occurring deep within them. These wavelengths provide detailed profiles of molecular transitions associated with dust. Because of its importance, astronomical observations centered on this wavelength region will chart star and planet formation in the Milky Way, the galactic life cycle of the elements, and the molecular and dusty universe.

Mid-IR spectra of individual objects such as H II regions, reflection nebulae, and planetary nebulae as well as the general interstellar medium are dominated by a set of emission features due to large aromatic molecules (a.k.a "PAH bands"). Studies of the spectral characteristics of such molecules and their dependence on molecular structure and charge state are of key importance for our understanding of this ubiquitous molecular component of the ISM. At long wavelengths, the continuum dust opacity is uncertain by an order of magnitude. IR spectral features of interstellar dust grains are used to determine their specific mineral composition, hence their opacities, which determine inferred grain temperatures and the masses of dusty objects, including the interstellar medium of entire galaxies. Emission bands from warm astronomical environments such as circumstellar regions, planetary nebulae, and star-forming clouds lead to the determination of the composition and physical conditions in regions where stars and planets form. The laboratory data essential for investigations of dust include measurements of the optical properties of candidate grain materials (including carbonaceous and silicate materials, as well as metallic carbides, sulfides, and oxides) as a function of temperature. For abundant materials (e.g., forms of carbon such as PAHs), the measurements should range from gas-phase molecules to nanoparticles to bulk materials. The IR spectral region is critical for the identification of grain composition, but results are also required for shorter wavelengths (i.e., UV), which heat the grains. The UV spectral region contains features of important large interstellar molecules, such as organic species which carry the IR emission bands and diffuse interstellar bands (DIBs) and that may be related to the origin of life. Studies of the UV characteristics of such molecules and their dependence on molecular structure and charge state are of importance for our understanding of this molecular component of the ISM. The lack of experimental data in this spectral region has hampered progress in theoretical studies as well as the interpretation of astronomical data. UV spectra are uniquely capable of identifying specific molecules, in contrast with the less specific transitions observed in the IR. Laboratory studies provide spectroscopy of large organic molecules and their ions. This work must be complemented by quantum theory calculations so that the laboratory data are properly interpreted.

### 4.4. Plasma Physics

Much of the Galactic neighborhood is in the plasma state. Modeling of plasma systems is beset with difficulties - none of turbulence, radiative energy transfer, collisionless shocks, and complex magnetic geometries can now be accurately modeled over scales of interest. As a result, laboratory plasma physics has much to contribute to our understanding of the corresponding dynamics.

Plasma hydrodynamic experiments can explore well-scaled dynamics of systems of interest, building on recent experience in which an experiment (Klein et al. 2003) has been used (Hwang et al. 2005) to directly interpret Chandra results observing a clump being destroyed by an SNR blast wave. This will contribute to the interpretation of data on the interactions of blast waves, shocks, and winds with interstellar structure from future observatories. Laboratory environments can produce radiating systems having dimensionless parameters in regimes relevant for example to shock breakout from supernovae, which will be relevant to the anticipated great increase in such data and to the simulation codes used to interpret it. To benchmark the models that are needed to interpret data from the photoionized regions mentioned above, one can irradiate a plasma volume containing relevant species with an intense x-ray source (e.g., Foord et al. 2006). In accretion disks, the magnetorotational instability (MRI) may provide the needed angular momentum transport. The MRI will be produced and studied in the laboratory during the next decade. This will prove to be of value in benchmarking the newly complex MHD codes needed to better understand phenomena such as stellar outflows, supernovae, and accretion disks. Other experiments that will contribute to such understanding include those on magnetic reconnection, which is already very active (e.g., Ren et al. 2005), on turbulent dynamos, which is poised to become active, and on collisionless shocks, which is feasible but requires progress with experimental facilities.

### 4.5. Nuclear Physics

Major uncertainties remain in our understanding of the nuclear physics governing massive-star evolution. The most critical parameter in the hydrostatic evolution of supernova progenitors is the $^{12}C(\alpha,\gamma)$ cross section, which determines the relative masses of the C and O shells, and thus has a major influence on supernova nucleosynthesis. There are several experimental efforts underway to further constrain this cross section, which is complicated due to the contributions of sub-threshold resonances.

The path of the *r*-process is through heavy, neutron-rich nuclei unknown on Earth, but which form the equilibrium state of nuclear matter in such astrophysical explosions. One of the main goals of the Facility for Rare Ion Beams (FRIB), a planned new Department of Energy accelerator, will be to produce these isotopes for the first time, so that their masses and beta decay lifetimes can be measured. The goal will be to characterize the nuclear physics well enough that the nucleosynthetic output of supernovae can then be used to constrain aspects of the explosion, such as the dynamic timescale for the ejecta and the freeze-out radius – just as Big Bang nucleosynthesis has been used to constrain conditions in the early Universe. These studies will also provide clues to the nature of the first stars in the Galaxy. As for the *s*-process, there are uncertainties associated with long-lived, but unstable nuclei. Current plans involve harvesting isotopes with FRIB and providing these targets for use at other facilities. Additional progress is anticipated with the Spallation Neutron Source at Oak Ridge National Laboratory.

### 5. Four central questions and one area with unusual discovery potential

#### 5.1 Four central questions
- How have the abundances of chemical elements evolved over cosmic time?
- How are the Galactic disk and halo coupled through a dynamic interstellar medium?
- What is the relationship between the Galactic Center and active galactic nuclei throughout cosmic time?
- What is the life cycle of cosmic carbon and what are the chemical steps leading to life on Earth?

### 5.2 One area with unusual discovery potential
- The field of astrobiology holds the promise of discovering how life arose on Earth and how likely it is elsewhere in our Galactic neighborhood.

### 6. Postlude

Laboratory astrophysics and complementary theoretical calculations are part of the foundation for our understanding of GAN and will remain so for many generations to come. From the level of scientific conception to that of the scientific return, it is our understanding of the underlying processes that allows us to address fundamental questions in these areas. In this regard, laboratory astrophysics is much like detector and instrument development; these efforts are constantly necessary to maximize the scientific return from astronomical observations.